\documentclass{aa}  

\usepackage{graphicx}
\usepackage{txfonts}

\usepackage{orcidlink} %{0000-0002-2363-6568} my orcid (mbm)
\usepackage{multirow}
\usepackage[normalem]{ulem}
\usepackage{breqn}

\begin{document}

   \title{Low-ionization structures in planetary nebulae -- IV. The molecular hydrogen counterpart}

   %\subtitle{....}

   \author{M. B. Mari\orcidlink{0000-0002-2363-6568}
          \inst{1, }\inst{2}
          \and
          S. Akras\orcidlink{0000-0003-1351-7204}\inst{3}
          \and
          D. R. Gonçalves\inst{4}
          }

   \institute{Universidad Nacional de Córdoba, Observatorio Astronómico de Córdoba, Laprida 854, X5000BGR Córdoba, Argentina\\
              \email{mbmari@unc.edu.ar}
         \and
            Consejo Nacional de Investigaciones Científicas y Técnicas (CONICET), Godoy Cruz 2290, CABA, CPC 1425FQB, Argentina
         \and
             Institute for Astronomy, Astrophysics, Space Applications and Remote Sensing, National Observatory of Athens, Penteli GR 15236, Greece
         \and
            Observat\'orio do Valongo, Universidade Federal do Rio de Janeiro, Ladeira Pedro Antonio 43, Rio de Janeiro 20080-090, Brazil
             }

   \date{Received XXXXXX, XXXX; accepted XXXXXX, XXXX}

% \abstract{}{}{}{}{} 
% 5 {} token are mandatory
 
  \abstract
   {Low-ionization structures (LISs), found in all morphological types of planetary nebulae (PNe), are small-scale features prominent in emission from low-ionization species such as [N~{\sc ii}], [S~{\sc ii}], [O~{\sc ii}] and [O~{\sc i}]. Observational and theoretical efforts have aimed to better understand their origin and nature. Recently, the detection of molecular hydrogen (H$_2$) emission associated with LISs in a few PNe has added a new piece to the puzzle of understanding these nebular structures. 
   }
   {Although observational studies indicate that LISs are characterized by lower electron densities than their host PNe, model predictions suggest higher total densities in these structures. The detection of H$_2$ emission from LISs in more PNe could help reconcile the observations with model predictions.
   }
   {Observations of five PNe with already known LISs were conducted using the Near InfraRed Imager and Spectrometer (NIRI) mounted on the 8~m Gemini North telescope. A narrow band filter, centered on H$_2$~1-0 2.122~$\mu$m emission line, was used along with a continuum filter, to ensure continuum subtraction.
   }
   {We present a deep, high-angular resolution near-IR narrowband H$_2$~1-0~S(1) imaging survey of five Galactic PNe with LISs. We nearly double the sample of LISs detected in the H$_2$ 1-0 2.122~$\mu$m emission line as well as the number of host PNe. These findings allows us to prove that the systematically lower electron density in LISs --~relative to the rims and shells of their host nebulae~-- is linked to the presence of H$_2$ molecular gas. Additionally, we provide the first estimation of the excited H$_2$ molecular mass in LISs, which is found between 200 and 5000 times lower than the corresponding ionized gas mass.
   
   }
   {}


   \maketitle
%
%-------------------------------------------------------------------

\section{Introduction}

As a low- or intermediate-mass star consumes its fuels, and it approaches the end of the evolutionary journey of its life, it ejects the outer layers of material into the interstellar medium, leaving behind a hot and luminous core. The radiation field of the exposed core peaks in the ultraviolet (UV) range, resulting in the dissociation of molecular gas formed in the previous stage of asymptotic giant branch (AGB) star and then in the ionization of atomic gas. 
These processes form the bright, and colourful planetary nebulae (PNe) we observe, 
which are among the most important contributors to the chemical enrichment of the interstellar medium in galaxies.

PNe display a variety of shapes and morphologies like round, elliptical, bipolar and multipolar \citep[e.g.][]{1996ApJ...466L..95M}. They are also characterized by distinct components such as rims, shells, and attached haloes prominent mostly in [{\sc O~iii}] and hydrogen recombination lines \citep[][]{2002ARA&A..40..439B}.
Besides, it is also known that PNe display structures in smaller scales prominent in emission from low-ionization species such as N$^{+}$, S$^{+}$, O$^{+}$ or O$^{0}$ \citep[e.g.][]{1996A&A...313..913C,1998AJ....116..360B,2001ApJ...547..302G}.

Many efforts have been made to decipher the nature of these low-ionization structures (hereafter LISs) and their unforeseen low electron density relative to the surrounding gas \citep[see e.g.][]{1994ApJ...424..800B,1997ApJ...487..304H,2003ApJ...597..975G,2004MNRAS.355...37G,2009MNRAS.398.2166G,2016MNRAS.455..930A,2016AJ....151...38D,2017PASA...34...36A,2020A.A...634A..47M,2021arXiv210505186M,2022MNRAS.512.2202A,2023MNRAS.518.3908M}. Recently, \citet[][]{2023MNRAS.525.1998M} probed the physico-chemical properties of LISs and host PNe, conducting a statistical analysis for a sample of 33 PNe containing 88 rims/shells and 104 LISs, and verified that neither the chemical abundances nor the electron temperatures show significant differences from one to another component. However, LISs correspond to a group of nebular structures statistically different from the rims and attached shells of the host nebula in terms of electron density ($n_e$), having two-thirds lower $n_e$ ($\sim$1700~cm$^{-3}$) than other nebular components ($\sim$2700~cm$^{-3}$), with interquartile ranges (25th to 75th percentiles) of 800-2700 cm$^{-3}$ and 1800–5000 cm$^{-3}$, respectively.

\citet{Odell1996} discussed the origin of the cometary knots in the well-studied Helix Nebula, suggesting that Rayleigh–Taylor instabilities during the early PN phase are the most likely cause, although a primordial origin linked to the central star’s formation cannot be ruled out. Other authors have also suggested similar formation mechanisms to explain the origin of LISs, such as stagnation points \citep{2001ApJ...556..823S}, Rayleigh-Taylor instabilities \citep[e.g.][]{Ramos_Phillips_2009}, AGB fossils \citep[e.g.][]{2001ApJ...547..302G}, among others. All these models require LIS' densities orders of magnitude higher than the surrounding nebular gas, typically $>$10$^4$~cm$^{-3}$ \citep[e.g.,][]{2008A&A...489.1141R,2020ApJ...889...13B}. \citet{2009MNRAS.398.2166G}, in an attempt to reconcile the optical observations with theoretical models, proposed that a significant fraction of gas in LISs should be neutral (atomic and/or molecular) and thus hidden for the visible light observational studies.

The high total (i.e., electron, neutral and molecular) density is a necessary condition for preventing molecular hydrogen (H$_{2}$) dissociation. The presence of H$_{2}$ in PNe is already well known, particularly in bipolar PNe \citep[Gatley's rule, e.g.][]{Kastner1996}, mainly through observations of the ro-vibrational emission line H$_{2}$~v=1-0~S(1), centred at 2.122~$\mu$m. The Helix Nebula, one of the brightest and closest planetary nebulae to Earth, was among the first to provide evidence of H$_2$ associated with knotty structures, enabling comprehensive investigations of its cometary knots through both observational and modeling approaches \citep{Odell1996, Odell2000, LopezMartin2001,2005AJ....130.1784M,2006ApJ...652..426H,2009ApJ...700.1067M}. 

It was not long ago that the presence of H$_2$ was also verified in other microstructures embedded in PNe, as LISs.
In this context, it is worth mentioning that an empirical linear correlation between the [O~{\sc i}] $\lambda$6300 and H$_{2}$~1-0~S(1) emission-lines had already been demonstrated in early studies of PNe \citep{1988MNRAS.232..615R}. Following the link between these two lines, Akras and collaborators selected LISs with strong [O~{\sc i}] $\lambda$6300 emission and succesfully detected the H$_{2}$ 2.122$\mu$m line in several of them: K~4-47 and NGC~7662 \citep[][]{Akras2017}, as well as NGC~6543 and NGC~7009 \citep{Akras2020}. Two additional PNe with H$_{2}$-emitting knots have also been reported: Hu~1-2 \citep{2015MNRAS.452.2445F} and Hb~12 \citep[][]{2018ApJ...859...92F}.

\citet{Aleman2011} showed that H$_{2}$ is confined to a narrow region between the point where the ionized hydrogen fraction reaches 95\% and the outer boundary of the ionized region, where this fraction drops to 0.01\%. Within this transition zone, the peak of the emissivity of the H$_{2}$~2.122~$\mu$m line is found, along with those of low-ionization lines ([N II], [S II], [O I]) (see their Fig.~4), with spatial separations
that exceed the thickness of the ionization front.
More recently, \citet{2024A&A...689A..14A} presented the first spatially resolved MUSE image of the atomic carbon line [C~{\sc i}]~8727\AA\ in NGC~7009. This line originates only from the pairs of LISs suggesting the presence of high density, partially ionized gas. The same emission line has also been detected in the LISs of NGC 3242 
suggesting that it could be another common characteristic of these microstructures \citep{Konstantinou2025}. \citet{GarciaRojas2022} also presented the [C~{\sc i}]~8727\AA\ MUSE map for a number of PNe. Examining the radial stratification of emission lines in the K1 LIS of NGC~7009 \citep{2003ApJ...597..975G}, \citet{2024A&A...689A..14A} argued that photoevaporation of dense knots could explain the observations.

The ionization and photoevaporation of neutral clouds illuminated by the UV radiation of O and/or B stars were studied since long ago \citep{Kahn1954,Oort1955}, aiming to explain their acceleration. Years later, \cite{Bertoldi1989,Bertoldi1990} developed approximate analytical models for the evolution of neutral clouds. The photoevaporation of the cloud's gas results in the rocket effect causing the clouds to accelerate away from the ionizing source. In such neutral/molecular clouds, the ionization and photodissociation fronts are not necessarily separated, which is a key difference from classical photodissociation regions (PDRs) \citep{Bertoldi1996,Henney2007}. Subsequently, a number of theoretical and observational studies applied the photoevaporation mechanism to neutral clumps in PNe \citep[e.g.][]{Mellema1998, Raga2005}, and to the cometary knots in Helix nebula \citep[e.g.][]{Odell2000,LopezMartin2001, Odell2005}. Nevertheless, LISs in PNe that do not show cometary morphology remain poorly understood, and the H$_2$ molecular component has only been observationally confirmed in a handful cases. 

This pilot survey presents the H$_2$ emission line of a sample of LISs in PNe, 
covering a variety of LIS' types in PNe of different morphological classes. The main goal of this study is analyzing and quantifying the molecular H$_2$ gas content of LISs. The detection of H$_2$ emission exclusively associated with LISs also enables a comparison between the masses of ionized and molecular material. Such an analysis can provide a more accurate estimate of the amount of molecular mass, at least in its excited state, within the LISs.
Through this approach, we aim to advance the understanding of a key discrepancy: while theoretical models predict significantly higher total densities in LISs compared to the rims/shells of their host PNe, optical observations indicate that the electron densities in LISs are at least two-thirds lower than those of the surrounding gas. This study will contribute to diminishing the gap between models and observations, providing crucial data to resolve this intriguing dichotomy.

In Section~\ref{sec2}, we present the observed sample along with the data reduction and calibration. In Section~\ref{sec3}, we show the results obtained from NIRI images. Section~\ref{sec4} provides the mass estimates and the relationship between ionized and excited molecular content. Finally, in Section~\ref{sec5} and \ref{sec6}, we present the discussion and overall conclusions.

%--------------------------------------------------------------------
\section{Narrowband imaging observations}\label{sec2}
Aiming to prove the hypothesis that the presence of H$_2$ is a general characteristic of LISs and is 
therefore associated with different LISs classes --~knots, filaments, jets, whether in pairs or isolated~-- we conducted 
very deep narrowband H$_2$ 2.122~$\mu$m imaging
for a sample of five PNe. 
Observations were carried out with the Near-InfraRed Imager and Spectrometer (NIRI), mounted on the 8~m Gemini North telescope in Mauna Kea, Hawaii.

\subsection{Data acquisition and reduction}

The observations were conducted between July and October 2020 (Program ID: GN-2020B-Q-128, PI: S. Akras). The f/6 configuration, with a pixel scale of 0.117~arcsec and a FoV of 120~arcsec, was used since it is ideal for observing PNe with angular sizes smaller than 100~arcsec. The narrowband G0216 filter, with an effective width of $\sim327$~\AA, centered at 2.1239~$\mu$m, was used to isolate the H$_2$~1-0~S(1) line at 2.122~$\mu$m. For proper continuum emission subtraction, the G0217 filter, with an effective width of $\sim329$~\AA, centered at 2.0975~$\mu$m, was also employed. To increase the signal-to-noise ratio (S/N), several individual frames were obtained per object. 
The observations are detailed in Table~\ref{tab:log_H2}. In agreement with the Gemini baseline calibrations, GCAL flat frames and dark frames with different times based on the science images were obtained to correct thermal emission, dark current and/or hot pixels. Standard stars (SAO34401, GSPC~S813-D, FS~150 and FS~123) were also observed to flux calibrate the data.

\begin{table}[ht]
\caption{Observation log.}
\label{tab:log_H2}
\centering
\begin{tabular}{lccc}
\hline
\multirow{2}{*}{Object}       &  \multirow{2}{*}{Filter} & {\# Combined} & \multirow{2}{*}{Exp.time~[s]}   \\
& & {frames} & \\
\hline
\multirow{2}{*}{NGC~2392}     &  H$_2$~v=1-0~S(1)      &  12                &  240            \\
                              &  K-cont                &  6                 &  240            \\
\multirow{2}{*}{NGC~6751}     &  H$_2$~v=1-0~S(1)      &  12                &  180            \\
                              &  K-cont                &  8                 &  180            \\
\multirow{2}{*}{NGC~6818}     &  H$_2$~v=1-0~S(1)      &  12                &  180            \\
                              &  K-cont                &  7                 &  180            \\
\multirow{2}{*}{NGC~6884}     &  H$_2$~v=1-0~S(1)      &  11                &  200            \\
                              &  K-cont                &  6                 &  200            \\
\multirow{2}{*}{NGC~7354}     &  H$_2$~v=1-0~S(1)      &  11                &  200            \\
                              &  K-cont                &  7                 &  200            \\
\hline
\end{tabular}
\tablefoot{
The H$_2$~v=1-0~S(1) narrowband filter 
isolates the molecular H line centered at $\lambda_{c}=2.1239$~$\mu$m, whereas the K-cont filter ($\lambda_{c}=2.0975$~$\mu$m) is used to map the continuum emission, thus allowing the continuum subtraction of the 
H$_2$~v=1-0~S(1) images.}
\end{table}

Before starting data reduction, as recommended by Gemini\footnote{\url{http://www.gemini.edu/instrumentation/niri/data-reduction}}, the {\sc Python} routines CLEARIR.py and NIRLIN.py were used. The first one is needed to remove artefacts as vertical striping and/or horizontal banding, whereas the latter corrects the non-linearity of the detector.
The images were reduced using {\sc DRAGONS}\footnote{\url{https://dragons.readthedocs.io/en/stable/}} \citep[Data Reduction for Astronomy from Gemini Observatory North and South, ][]{2019ASPC..523..321L}, a {\sc Python}-based software for the Gemini data reduction, provided by the Gemini observatory. This software package can stack multiple dark observations to create a master dark. From the GCAL flats and short darks, a bad pixel mask (BPM) can be produced. A master flat can also be generated from multiple observations of lamps-on and lamps-off. Science data were then reduced considering the BPM, the corresponding master dark 
and the appropriate master flat field for each filter. 

\subsection{Flux calibration and continuum subtraction}
\label{sec-fluxcalib}
Given the night-time atmospheric variations, the 
full width at half maximum (FWHM) of the background stars was not necessarily the same in both filters. Using 
background stars present in both images, we computed, per target, the average FWHM. 
The image with the smallest FWHM was degraded before proceeding with the subtraction of the continuum from the emission-line image.
This was done employing the {\sc Gauss} task in {\sc IRAF}, which convolves the image with an elliptical Gaussian function after specifying the sigma parameter.

Due to the lack of broadband K filter observations for the standard stars, we calculated their theoretical K magnitudes (Flux$_{K~theoretical}$) by modelling their blackbody spectral energy distributions (SEDs). 
For this end, we adopted effective temperatures derived from their spectral types and used the transmission curves of the filters. For flux calibration, we used the absolute K magnitudes of the standard stars (Flux$_{K~standard~ star}$) from the Two Micron All Sky Survey \citep[2MASS;][]{2006AJ....131.1163S} for SAO34401 and from the UKIRT Infrared Deep Sky Survey \citep[UKIDSS;][]{2008MNRAS.384..637H} for GSPC~S813-D, FS~150, and FS~123. From this, we derived the calibration factors (Flux$_{K~standard~ star}$/Flux$_{K~theoretical}$), which were then used to flux-calibrate the narrowband images of the targets obtained with NIRI, applying the transmission curves of the corresponding filters.

\begin{figure}[ht]
    \centering
    \includegraphics[width=0.32\columnwidth]{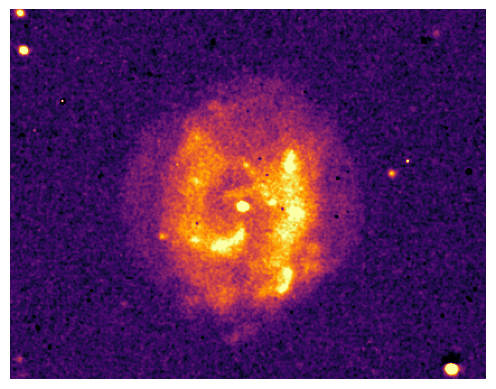}
    \includegraphics[width=0.32\columnwidth]{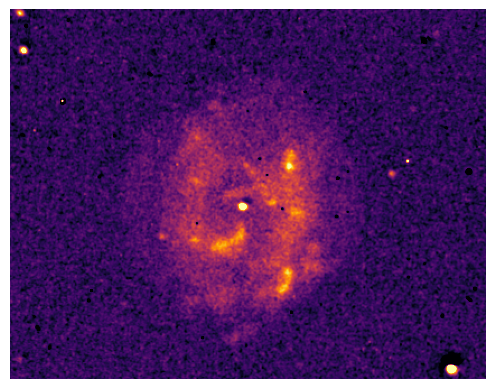}
    \includegraphics[width=0.32\columnwidth]{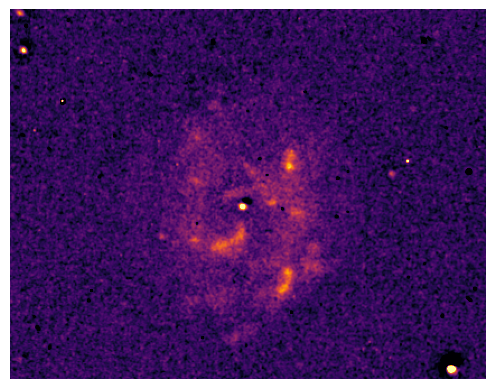}
    \includegraphics[width=0.32\columnwidth]{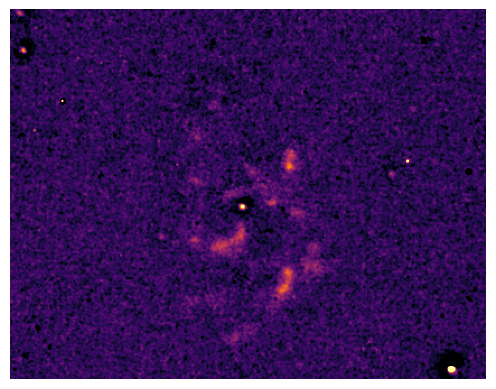}
    \includegraphics[width=0.32\columnwidth]{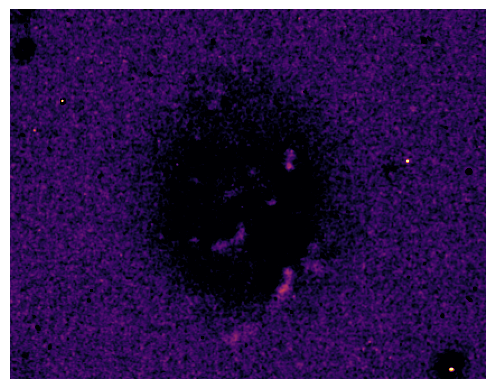}
    \includegraphics[width=0.32\columnwidth]{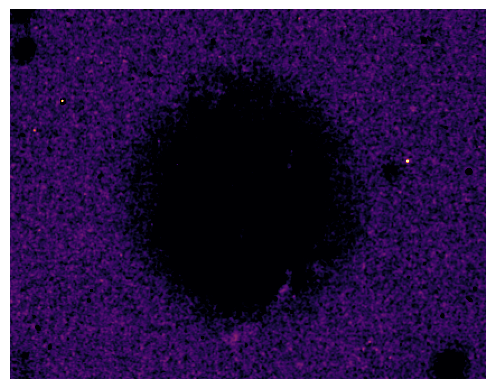}
    \caption{ Comparison of continuum-subtracted images that resulted from a variation of 
    scale factors. From top to bottom and left to right, the scale factor increases (0.9, 1.0, 1.05, 1.1, 1.2, 1.3). Initially, 
    the poor continuum subtraction is evident, since the field stars appear as H2 line emitters (top panels). The last two bottom panels,  are clear examples of  over-subtraction of the continuum.}  
    \label{fig:comparation}
\end{figure}

For continuum subtraction in the flux-calibrated images, we first aligned the H$_2$ and K-cont images, using as many field stars as possible. Then, we computed the scale factor required for the proper subtraction of field stars in the emission-line frames. Specifically, we used the equation $filter_{H_{2}}-(H_{2}/K{\sc -cont}) \times filter_{K-cont}$, where 
(H$_{2}/$K--cont) represents the scale factor. To estimate this factor, background stars present in both images were used, and their fluxes 
were compared. Fig.\ref{fig:comparation} gives an example of the continuum-subtraction process, for which we apply different scale factors. A scale factor that is too small (insufficient) is imperceptible in the continuum subtraction, while excessively high values lead to oversubtraction, resulting in an underestimation of the H$_2$ fluxes.

Several uncertainties are involved in the process just described, including those due to: (i)~the standard stars effective temperature used to obtain their theoretical SEDs; (ii)~the conversion factors applied for the flux calibration of both standard stars and science targets; and (iii)~the scale factor used for continuum subtraction, determined by ensuring satisfactory removal of field stars in the H$_2$ science images. 
Additionally, flux calibration is subject to an inherent photometric uncertainty of approximately 10\% due to unmeasured atmospheric extinction\footnote{\url{https://www.gemini.edu/observing/resources/near-ir-resources}}. Each of the quoted processes contributes with an uncertainty of roughly 10\%, while these uncertainties may not be entirely independent. Moreover, both 2MASS and UKIRT report photometric errors of approximately 0.02-0.03~mag \citep[][, respectively]{2006AJ....131.1163S,2007MNRAS.379.1599L}, resulting in a baseline flux calibration uncertainty of around $\sim$3\%. The combined effect of these uncertainties likely results in an overall uncertainty in the H$_2$ flux measurements of about 30-40\%. 

Despite these uncertainties, we anticipate that when comparing the parameter A in the relation M$_{H~II}$/M$_{H_{2}}$ = A/$n_{e}\times$F$_{Br\gamma}$/F$_{H_{2}}$ with values reported in the literature, we find a difference of only 13\% (see Section~\ref{sec4.1} where we define and discuss the H~{\sc ii} over H$_{2}$ mass ratio). This level of agreement suggests that, while uncertainties are significant, our flux calibration methodology produces reliable results.
 
%--------------------------------------------------------------------
\section{Results: NIRI H$_2$ versus HST [N~{\sc ii}] imagery}\label{sec3}

In this work, the first deep, spatially resolved and continuum-subtracted narrowband H$_2$ 2.122~$\mu$m images for the planetary nebulae NGC~2392, NGC~6751, NGC 6818, NGC~6884 and NGC~7354 are presented. 
The NIRI H$_2$ results are presented in comparison with HST [N~{\sc ii}] images of this PNe sample, which facilitates the interpretation of our results.

\vspace{0.3cm}

\subsection{NGC~2392}

\begin{figure*}[ht!]
    \centering
    \includegraphics[width=0.7\textwidth]{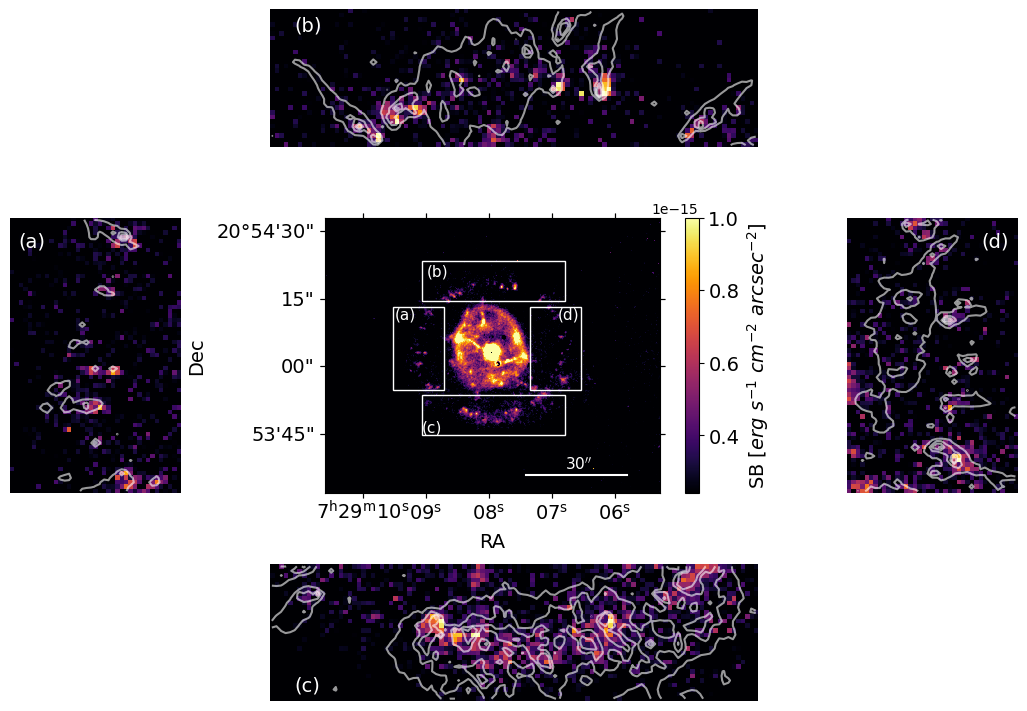}
    \caption{Surface brightness (SB) of the H$_{2}$~1–0~S(1) continuum-subtracted image for NGC 2392 (center). Panels a, b, c, and d display the H$_{2}$ continuum-subtracted image overlaid with HST [{\sc N~ii}] emission contours for different regions of the nebula.}
    \label{fig:n2392}
\end{figure*}

It is known that in addition to the multiple system of knotty and filamentary LISs, located at the inner part of its high-excitation shell, NGC~2392 also has a bipolar jet of high-velocity (150-185~km~s$^{-1}$), directed almost toward the observer \citep{1985ApJ...289..526O,1985ApJ...295L..17G}. Due to their orientation and low surface brightness, these jets are not detected in optical images and can only be distinguished in spatially resolved high-resolution [N~{\sc ii}] spectra \citep[see e.g.][]{2012ApJ...761..172G,2021ApJ...909...44G}. In the central panel of Fig.~\ref{fig:n2392}, we present the narrowband continuum-subtracted H$_2$ 2.122~$\mu$m image of NGC~2392. Multiple knots and filaments are seen in the molecular hydrogen light. Panels (a), (b), (c) and (d) cover  
different regions of the nebula, and also display the contours of the [N~{\sc ii}]~$\lambda$6584 emission from HST, overlaid on the NIRI H$_2$~1–0~S(1) image NGC~2392. H$_2$ and [N{\sc~ii}] lines show a very good spatial correlation, as it has been demonstrated in previous studies \citep{Akras2017,Akras2020}. The H$_2$ fluxes from five LISs are listed in Table~\ref{tab:masses}, 
they range from 0.25 to 0.8 $\times$ 10$^{-15}$ erg s$^{-1}$cm$^{-2}$.\\

\subsection{NGC 6751}

\begin{figure*}[ht!]
    \centering
    \includegraphics[width=0.7\textwidth]{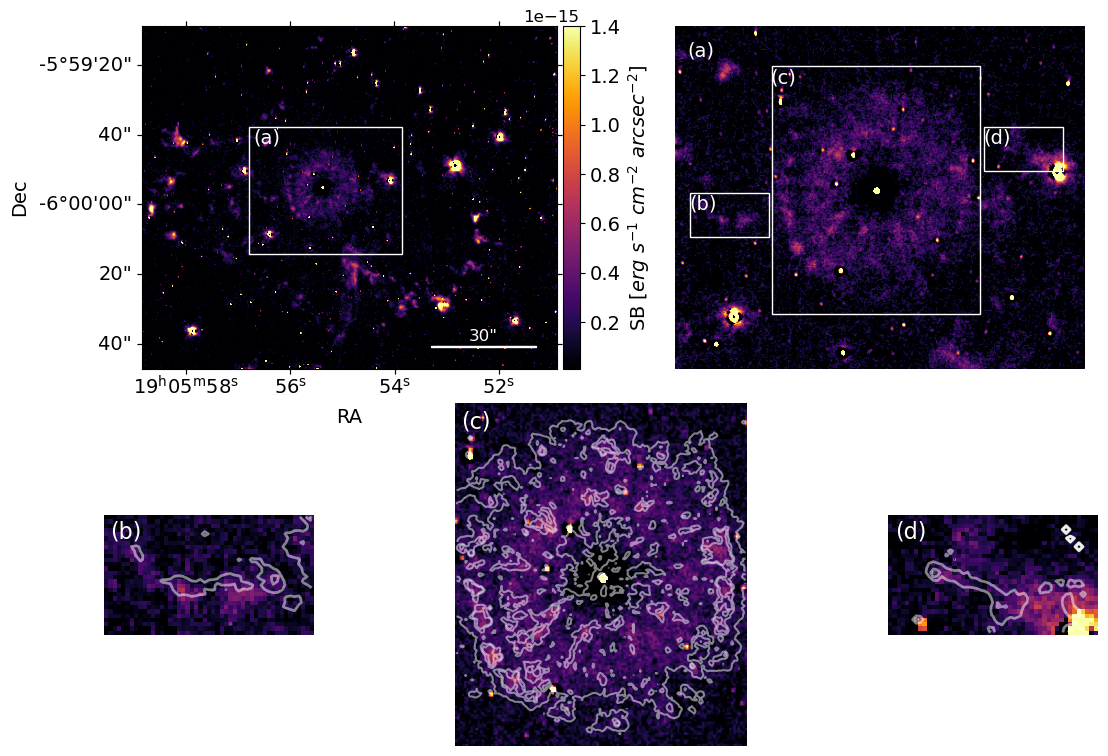}
    \caption{Surface brightness (SB) of the H$_{2}$~1–0~S(1) continuum-subtracted image for NGC~6751. The upper-left panel displays the entire nebula, revealing faint emission from its halo at distances greater than 30~arcsec. The right panel (a) provides a zoomed-in view of the nebula's center, showing the eastern and western jet-like structures along with numerous knots in the outer rim. The bottom panels present the continuum-subtracted H$_{2}$ image overlaid by the HST [{\sc N~ii}] emission line contours: (c) focuses on the nebula's center, while (b) and (d) display the pairs of the jet-like structures. The seemingly brilliant knot in the western jet-like structure is a background star. }
    \label{fig:n6751}
\end{figure*}

Among its multitude of structures, this PN has a [N~{\sc ii}] knotty ring and irregular filaments in the northeast quadrant of the halo \citep{1991ApJ...376..150C,2010ApJ...722.1260C}. Figure~\ref{fig:n6751} presents the continuum-subtracted H$_2$ images of NGC~6751.
It can be seen that H$_2$ emission emanates from the inner parts of the nebula, where [N{\sc~ii}] emission is evident.  Faint H$_2$ emission is 
detected in the pairs of jet-like structures which are also detected in the [N{\sc~ii}] line. The H$_2$ fluxes of the latter structures vary from 0.37 to 1.73 $\times$ 10$^{-15}$~erg~s$^{-1}$cm$^{-2}$ (see Table~\ref{tab:masses}). The filamentary structures of the 
halo exhibit stronger H$_2$ emission (upper left panel), with fluxes ranging from 1.85 to 3.84~$\times$~10$^{-15}$~erg~s$^{-1}$cm$^{-2}$. These findings further supports the \citet{2010ApJ...722.1260C} idea that such filaments are remnants of AGB mass-loss events. 

\subsection{NGC~6818}

\begin{figure*}[ht!]
    \centering
    \includegraphics[width=0.75\textwidth]{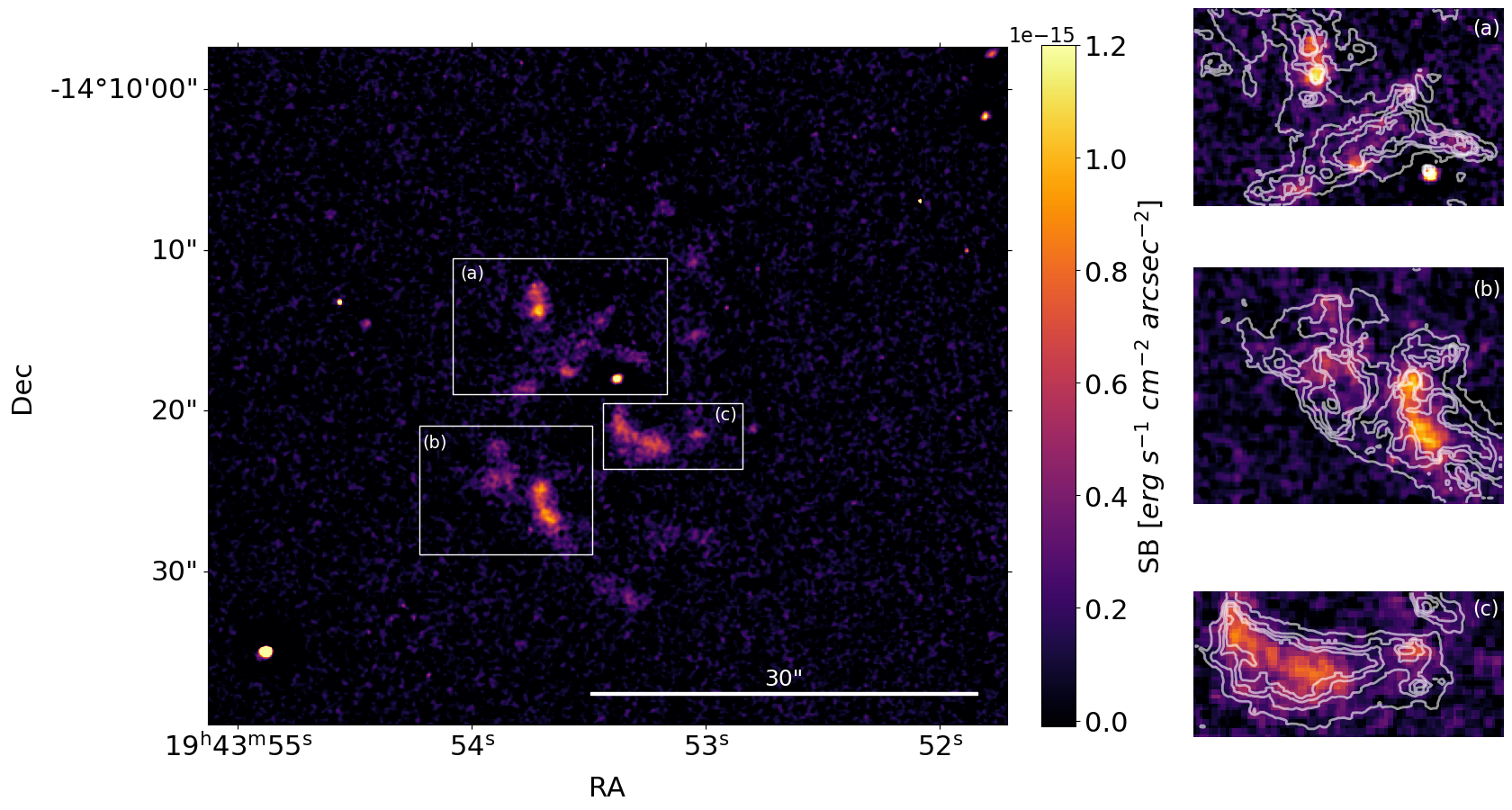}
    \caption{Surface brightness (SB) of the H$_{2}$~1–0~S(1) continuum-subtracted image for NGC~6818 is shown in left panel. The other panels present the H$_{2}$ continuum-subtracted image overlaid with HST [{\sc N~ii}] emission contours, highlighting three regions of the nebula: (a) and (b) correspond to the "moustache" structures, while (c) is the brighter southern part of the nebula.} 
    \label{fig:n6818}
\end{figure*}

In the light of [N~{\sc ii}], NGC~6818 shows multiple filamentary structures distributed around the whole nebula, labeled as "moustaches" by \citet{2003A&A...400..161B}. Our deep near-IR image of this nebula is presented in Fig.~\ref{fig:n6818} along with the [N~{\sc ii}] emission line contours from HST. H$_{2}$ 2.122~$\mu$m emission line is detected only in the LISs or "moustaches", 
prominent in [N~{\sc ii}]. The H$_2$ flux based on three LISs varies between 1.44 and 2.88 $\times$ 10$^{-15}$ erg s$^{-1}$~cm$^{-2}$ (see Table~\ref{tab:masses}), which are among the brightest LISs in our sample.

\subsection{NGC~6884}

\begin{figure*}[ht!]
    \centering
    \includegraphics[width=0.85\textwidth]{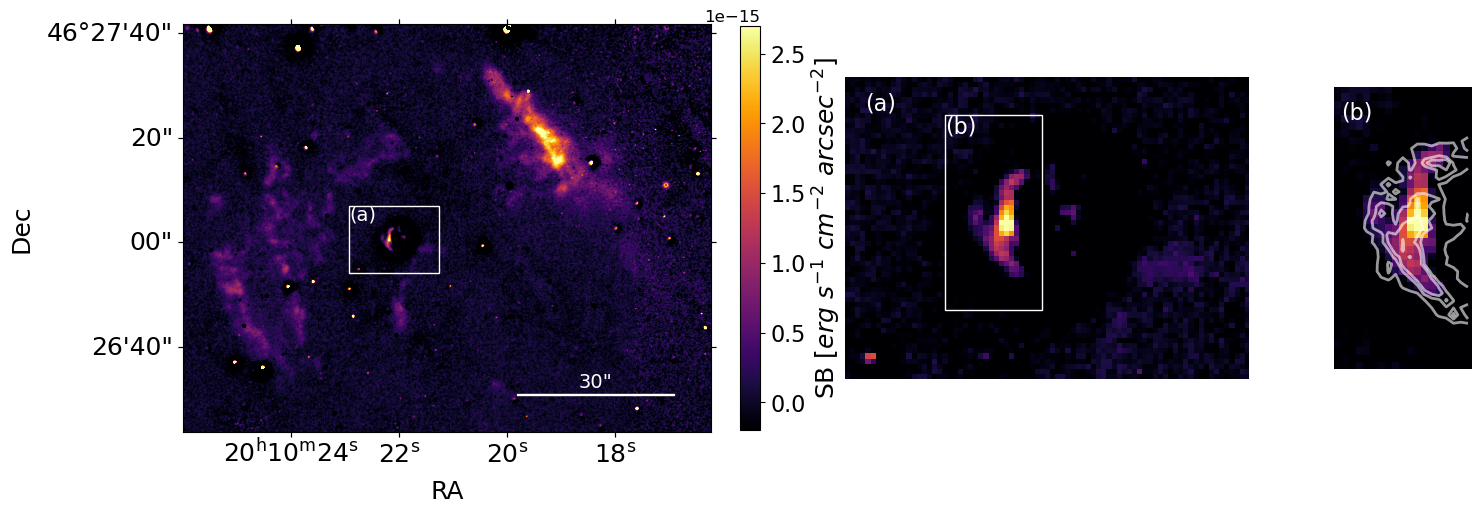}
    \caption{Surface brightness (SB) of the H$_{2}$~1–0~S(1) continuum-subtracted image for NGC~6884 is shown in left panel. The center panel provides a zoomed-in view of the entire nebula, while the other two panels display the H$_{2}$ continuum-subtracted image overlaid with HST [{\sc N~ii}] emission contours, highlighting the knotty arc-like structure in a half region of the nebula, labeled as (b) and (c). } 
    \label{fig:n6884}
\end{figure*}

The [N~{\sc ii}] image of NGC~6884 shows a pair of LISs in opposite direction with a radial velocity of 40~km~s$^{-1}$ \citep{1999AJ....117.1421M}, interpreted at that time as a precessing bipolar outflows. 
More recent, diffraction-limited and narrow-band HST images in the [N~{\sc ii}] emission 
unveiled that the pair of knots is actually two symmetrical knotty arc-like structures \citep{2002AJ....123.2666P}.
Figure~\ref{fig:n6884} shows the continuum-subtracted H$_2$ 2.122~$\mu$m image of NGC~6884. 
In the left panel, we can observe that significant H$_2$ is detected for the time from the outer halo of this nebula, at distances up to 30-40 arcsec from the central star. Panels (a) and (b) present a zoom-in view of the central part of the nebula, where an arc-like structure is clearly detected in the eastern part. The H$_{2}$~(1-0)~S(1) emission of this arc-like structure shows a spatial correlation with the [N~{\sc ii}] emission from the HST data, but we cannot verify that H$_2$ emission originates from the knots due to the lower-spatial resolution of our data. 
The H$_2$ flux in two regions of the eastern arc-like ranges from 0.75 to 1.86$\times$ 10$^{-15}$~erg~s$^{-1}$~cm$^{-2}$ (see Table~\ref{tab:masses}), while regions in the outer halo exhibit fluxes between 2.60 and 18.35~$\times$~10$^{-15}$~erg~s$^{-1}$cm$^{-2}$. As in NGC~6751, the H$_2$ emission from halo structures in this object also seems to be related to remnants of mass-loss episodes during the AGB stage.

\subsection{NGC~7354}

\begin{figure*}[ht!]
    \centering
    \includegraphics[width=0.69\textwidth]{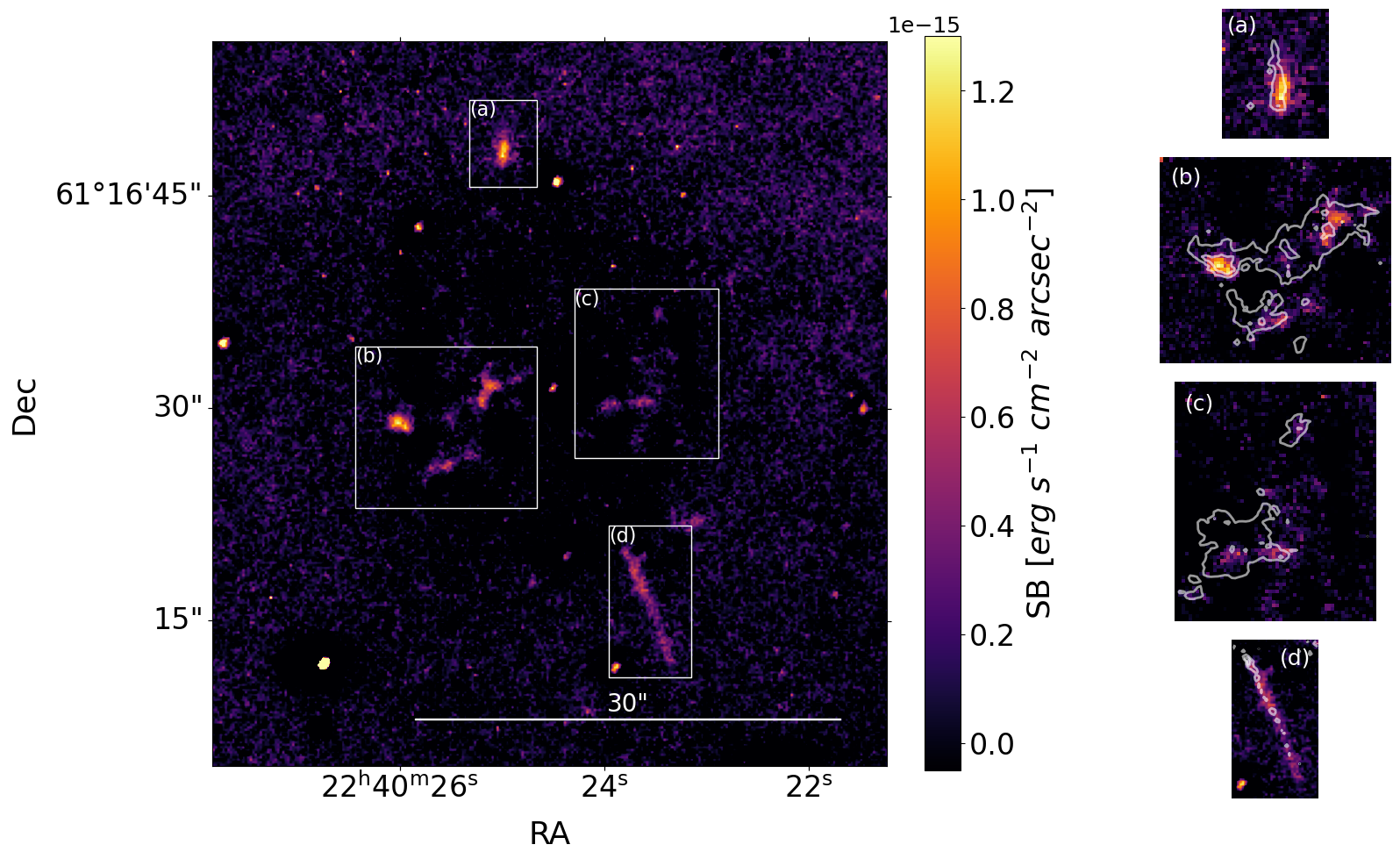}
    \caption{Surface brightness (SB) of the H$_{2}$~1–0~S(1) continuum-subtracted image for NGC~7354 is shown in left panel. The other panels display the H$_{2}$ continuum-subtracted image overlaid with HST [{\sc N~ii}] emission contours, highlighting for four regions of the nebula: (a) and (d) correspond to the pair of jet-like structures, while (b) and (c) show the equatorial bright LISs.} 
    \label{fig:n7354}
\end{figure*}

Multiple microstructures prominent in [N~{\sc ii}] emission have been identified and studied by a few authors in NGC~7354 \citep[e.g., ][]{1997ApJ...487..304H,2010AJ....139.1426C}, including the pair of jet-like LISs along its major axis. By using near- and mid-IR data from 2MASS and \textit{Spitzer}, \citet{2009MNRAS.399.1126P} studied NGC~7354. Due to the low spatial resolution of their data, the  equatorial LISs and jet--like structures were not clearly detected.  Nevertheless, all LISs are seen in our data (Fig.~\ref{fig:n7354}). The left panel presents the continuum-subtracted H$_2$ image of NGC~7354. The right panels (a), (b), (c) and (d) show the contours of [N~{\sc ii}] emission superimposed on the H$_2$ zoom-in images of the LISs. There is a clear spatial correlation between the two emission lines. We also report the first detection of H$_2$ emission from the jet-like structures indicative of collisional excitation. The H$_2$ fluxes of four LISs vary from 1.35 to 2.68 $\times$ 10$^{-15}$ erg s$^{-1}$~cm$^{-2}$ (see Table~\ref{tab:masses}).

%--------------------------------------------------------------------
\section{Results: molecular and ionized LISs' content}\label{sec4}

The primary motivation for studying LISs' emission in molecular hydrogen lines was to address the question: {\it Why do the vast majority of LISs exhibit electron densities lower than those of nebular rims and shells?} Since the latter contradicts theoretical predictions, we computed the molecular and ionized masses in LISs to examine whether their H$_2$ content helps alleviate the tension between observations and theory, as previously suggested by \citet{2009MNRAS.398.2166G}. Here, we integrate our results with data available in the literature.

\vspace{0.3cm}

\subsection{Molecular hydrogen mass}

After measuring the H$_2$~$\lambda$2.122 fluxes for several LISs in our sample, 
we can also estimate their molecular hydrogen mass. 
For that we adopted Equation~\ref{eq1} from \citet{1982ApJ...253..136S}:
\begin{dmath}\label{eq1}
F_{H_{2}\lambda2.122} = \frac{n_{H_{2}}V_{H_{2}}f_{\nu=1,J=3}A_{S(1)}h\nu}{4\pi d^{2}}
\end{dmath}

\noindent where n$_{H_{2}}$ is the H$_2$ volume density, V$_{H_{2}}$ is the volume of the region that contains the 
H$_2$ gas, A$_{S(1)}= 3.47 \times 10^{-7} s^{-1}$ is the H$_2$~1-0~S(1) transition probability for temperature T=2000~K \citep{1977ApJS...35..281T,2008MNRAS.385.1129R}, $f_{\nu=1,J=3} = 0.0122$ is the population fraction of H$_2$ in the $\nu$=1, J=3 level, \textit{d} is the distance of the nebula, \textit{h} is the Planck constant, and $\nu$ is the frequency of the H$_2$ line. 
Rearranging the above equation allows us to determine the mass of the H$_2$ component in LISs, which can be derived from:

\begin{dmath}\label{eq2}
  M_{H_{2}} = \frac{2m_{p}F_{H_{2}\lambda2.122}4{\pi}d^{2}}{f_{\nu=1,J=3}A_{S(1)}
h\nu} \sim 5.0776\times10^{7}\left(\frac{F_{H_{2}\lambda2.122}}{erg~s^{-1}
cm^{-2}}  \right)\left( \frac{d}{kpc} \right)^{2} [M_{\odot}]
\end{dmath}

\noindent where m$_{p}$ is the mass of the proton and F$_{H_{2}~\lambda2.122}$ is the observed flux of the emission line \citep{2015MNRAS.453.1727D}.

Since it requires an external energy source to excite the ro-vibrational levels of the H$_2$ molecule, such as UV photons from the central star or shock waves, the flux from the H$_2$~$\lambda$2.122 line does not represent the distribution of the cold molecular component. Consequently, our data 
trace the highly excited H$_2$ gas, referred to as the \textit{warm} H$_2$ component, and therefore the masses estimated from the aforementioned equation correspond to this component. 
Thus, the masses computed and presented in this work represent 
the lower limits of the total H$_2$ mass in LISs.

Table~\ref{tab:masses} encompasses the first estimations of the \textit{warm} H$_2$ masses for 19 LISs in our sample of five PNe. 
These masses range from 0.4 to 10$\times$10$^{-7}$~M$_{\odot}$, with an average of 4.6$\times$10$^{-7}$~M$_{\odot}$. 
Given that Equation~\ref{eq2} 
is commonly applied to active galactic nuclei \citep[AGN; e.g., ][]{2002MNRAS.331..154R,2015MNRAS.453.1727D}, 
we also computed the H$_2$ mass of the cometary knot K1 (RA:22:29:33.41, Dec:-20:48:04.73) in the Helix planetary nebula using the same equation to ensure the reliability of our estimates. 
This procedure resulted in a H$_2$ mass for K1 of $\sim$3.9$\times$10$^{-8}$~M$_{\odot}$ 
that is one order of magnitude 
smaller than the H$_2$ masses in the LISs of our PNe sample. Nevertheless, our measurement of the H$_2$ mass for K1 is 
reasonably in agreement with that provided by \citet{2007MNRAS.382.1447M} through near-IR IFU observations (see Table~\ref{tab:masses}). 
The detection of several H$_2$ lines from different ro-vibrational states allowed the authors to construct the excitation diagram and determine the excitation temperature and column density of the H$_2$ gas from a 
robust approach, resulting in 
$\sim2\times10^{-8}~M_{\odot}$, which is 
half of the value we obtained. Considering the different type of data, the uncertainties in the H$_2$ fluxes (discussed in Section~\ref{sec-fluxcalib}), 
and the fact we adopted the distance to the  nebula from Gaia DR3  parallaxes \citep{2023A&A...674A...1G,2021AJ....161..147B}, only available more recently, we argue that our methodology provides adequate and reliable warm H$_2$ masses for the LISs.

\citet{2024MNRAS.528.3392W} also provided an H$_2$ mass for the globules of NGC~6720, using data from the JWST Early Release Observations. The authors used the same approach as \citet{2022NatAs...6.1421D} 
for NGC~3132. 
The extinction of the knots in the IR wavelength regime was derived by assuming the extinction law 
from \citet{1989ApJ...345..245C}. This led first to the estimation of the column density, and 
then the total H density and H$_2$ mass. 
\citet{2024MNRAS.528.3392W} estimated 
densities of n$_{H}\sim10^{5}$~cm$^{-3}$ and
H$_2$ masses of $\sim10^{-6}$~M${_\odot}$. They noted that, at these densities ($n_{H_2}=0.5n_H$), the globules could be in pressure equilibrium with the surrounding ionized gas. Consequently, the globules could remain essentially stable, avoiding collapse or dissipation, until the surrounding gas recombines. On the other hand, \citet{2022NatAs...6.1421D} reported a higher density of n$_{H}\sim10^{6}$~cm$^{-3}$ and an H$_2$ mass of $\sim10^{-5}$~M${_\odot}$, for the filaments in NGC~3132.
These values are 2-3 orders of magnitude higher than the mass of the cometary knot in Helix and 1-2 orders of magnitude greater than our masses in LISs. Although an in-depth comparison of the different structures studied in the referenced works is beyond the scope of this study,
we argue that the methodology followed by \citet{2007MNRAS.382.1447M} provides robust estimates for densities and masses.

\begin{table*}[ht]
\caption{Position, H$_2$~1-0-S(1) fluxes, and molecular masses for various LISs in the studied PNe are presented. The distances required for these calculations were derived from the Gaia DR3 parallax measurements \citep{2023A&A...674A...1G}, yielding the following values: 1.83~kpc for NGC~2392, 3.44~kpc for NGC~6751, 2.86~kpc for NGC~6818 and 2.10~kpc for NGC~7354. For NGC~6884, the distance was adopted from \citet{2020ApJ...890...50G} and is 3.2~kpc.}
\label{tab:masses}
\centering
\small
\begin{tabular}{lllllcc}
\hline
Name                          & LISs      & RA               & Dec                 & Box                  & Flux H$_2$~1-0~S(1)                           & H$_2$ M$_{warm}$                                         \\
                              &           &                  &                     & [arcsec$^2$]         & [$\times 10^{-15}$ erg s$^{-1}$cm$^{-2}$]     &  [$\times 10^{-07}$ M$_{\odot}$]                          \\
\hline                              
\multirow{4}{*}{NGC 2392}     &  1     & 07:29:11.54      & $+$20:54:34.22        & 0.896$\times$0.478          & 0.25       & 0.4        \\
                              &  2     & 07:29:11.84      & $+$20:54:39.48        & 1.499$\times$0.483          & 0.38       & 0.6        \\
                              &  3     & 07:29:11.66      & $+$20:54:50.73        & 1.396$\times$0.551          & 0.37       & 0.6        \\
                              &  4     & 07:29:10.32      & $+$20:54:56.62        & 0.710$\times$1.464          & 0.80       & 1.4        \\
                              &  5     & 07:29:09.90      & $+$20:54:53.93        & 0.539$\times$1.903          & 0.53       & 0.9        \\
\hline                              
\multirow{9}{*}{NGC 6751}     &  1     & 19:05:56.60      & $-$05:59:36.90        & 1.273$\times$1.272          & 0.37       & 2.2        \\
                              &  2     & 19:05:56.46      & $-$05:59:37.21        & 1.992$\times$1.632          & 0.75       & 4.5        \\
                              &  3     & 19:05:56.08      & $-$05:59:39.95        & 1.692$\times$1.549          & 1.04       & 6.2        \\
                              &  4     & 19:05:55.80      & $-$05:59:37.22        & 0.873$\times$3.375          & 1.10       & 6.6        \\
                              &  5     & 19:05:55.80      & $-$05:59:29.84        & 1.832$\times$2.208          & 1.73       & 10.4        \\
                              &  H1$^{\mathsection}$    & 19:05:58.37      & $-$05:59:47.58        & 1.413$\times$2.194          & 2.26       & 13.6        \\
                              &  H2$^{\mathsection}$    & 19:05:52.57      & $-$05:59:42.75        & 1.317$\times$2.348          & 3.84       & 23.1        \\
                              &  H3$^{\mathsection}$    & 19:05:52.52      & $-$05:59:48.79        & 1.409$\times$2.584          & 2.27       & 13.6        \\
                              &  H4$^{\mathsection}$    & 19:05:58.41      & $-$05:59:32.30        & 1.351$\times$1.377          & 1.85       & 11.1        \\
\hline                              
\multirow{3}{*}{NGC 6818}     &  1     & 19:43:57.94      & $-$14:09:09.37        & 1.229$\times$2.517          & 1.44       & 6.0        \\
                              &  2     & 19:43:57.51      & $-$14:09:17.69        & 1.418$\times$4.050          & 2.39       & 10.0        \\
                              &  3     & 19:43:57.91      & $-$14:09:21.85        & 1.497$\times$2.761          & 2.17       & 9.0        \\
\hline                              
\multirow{6}{*}{NGC 6884}     &  1     & 20:10:23.72      & $+$46:27:37.68        & 0.545$\times$1.468          & 1.86       & 9.7        \\
                              &  2     & 20:10:23.75      & $+$46:27:36.39        & 0.616$\times$0.992          & 0.75       & 3.9        \\
                              &  H1$^{\mathsection}$    & 20:10:21.02      & $+$46:28:00.16        & 2.000$\times$2.000          & 6.67       & 34.7        \\
                              &  H2$^{\mathsection}$    & 20:10:20.68      & $+$46:27:55.67        & 3.000$\times$3.000          & 18.35       & 95.4        \\
                              &  H3$^{\mathsection}$    & 20:10:25.64      & $+$46:27:31.87        & 2.000$\times$2.000          & 2.60       & 13.5        \\
                              &  H4$^{\mathsection}$    & 20:10:26.57      & $+$46:27:28.04        & 3.000$\times$3.000          & 6.91       & 35.9        \\

\hline
\multirow{4}{*}{NGC 7354}     &  1     & 22:40:20.12      & $+$61:17:23.82        & 0.984$\times$2.349          & 1.35       & 3.0        \\
                              &  2     & 22:40:21.13      & $+$61:17:04.84        & 1.389$\times$1.784          & 1.41       & 3.1        \\
                              &  3     & 22:40:18.66      & $+$61:16:52.00        & 1.553$\times$10.70          & 2.69       & 6.0        \\
                              &  4     & 22:40:20.28      & $+$61:17:06.98        & 1.559$\times$2.940          & 1.47       & 3.3        \\
\hline                              
Helix  & \multirow{2}{*}{$^{\dag}$K1} & \multirow{2}{*}{$^{\dag}$22:29:33.41} & \multirow{2}{*}{$^{\dag}$-20:48:04.73} & \multirow{2}{*}{$^{\dag}$2x2} & \multirow{2}{*}{$^{\dag}$1.88$\times 10^{-14}$} & 3.9$\times 10^{-08}$              \\
Nebula &        &                              &                               &                      &                             & $^{\dag}$2$\times 10^{-08}$   \\
\hline 
\end{tabular}
\tablefoot{ The last row corresponds to another LIS, a cometary knot of Helix nebula, for comparison purposes. \\
$^{\mathsection}$ Structures found in the nebular halo. \\
$^{\dag}$The average intensity over 2$\times$2~arcsec$^{2}$ and the H$_2$ warm mass, are from \citet{2007MNRAS.382.1447M}. }
\end{table*}

\subsection{Ionized mass}\label{sec4.1}
In addition to the warm H$_2$ component, the ionized mass can also be estimated using near-IR observations, specifically from the flux of the Br$\gamma$ emission line:

\begin{dmath}\label{eq3}
    F_{Br\gamma} = 2.77\times 10^{-28} \frac{{n_{e}}^2V_{HII}}{d^{2}}.
\end{dmath}

\noindent In Equation~\ref{eq3}, assuming that the H~{\sc ii} region 
associated with the H$_2$ is optically thin in Br$\gamma$ \citep{1982ApJ...253..136S}, and adopting case B emissivities from \citet{2006agna.book.....O}, the ionized mass (M$_{HII}$) can be derived using the equation below:

\begin{dmath}\label{eq4}
   M_{HII} \sim 3.29\times10^{13}\left(\frac{F_{Br\gamma}}{erg~s^{-1}cm^{-2}}  
\right)\left( \frac{d}{kpc} \right)^{2} \left(\frac{n_{e}}{cm^{-3}}\right)^{-1} 
[M_{\odot}]
\end{dmath}
\noindent \citep{2009MNRAS.394.1148S}.

Unfortunately, the PNe in our sample were not observed in the Br$\gamma$ emission line. We then adopt the results in \citet{Akras2017,Akras2020}, who provides fluxes for a number of LISs in four PNe, both for the H$_2$~1-0~S(1) and Br$\gamma$ lines, ranging from 0.72 to 21.89$\times 10^{-15}$~erg~s$^{-1}$~cm$^{-2}$, and from 0.96 to 94.70$\times 10^{-15}$~erg~s$^{-1}$~cm$^{-2}$, respectively. In Table~\ref{tab:masses_akras}, we list the masses for the warm H$_2$ gas 
(obtained from Equation~\ref{eq2}), as well as the ionized mass 
(from Equation~\ref{eq4}). 

\begin{figure}[ht!]
    \centering
    \includegraphics[width=\columnwidth]{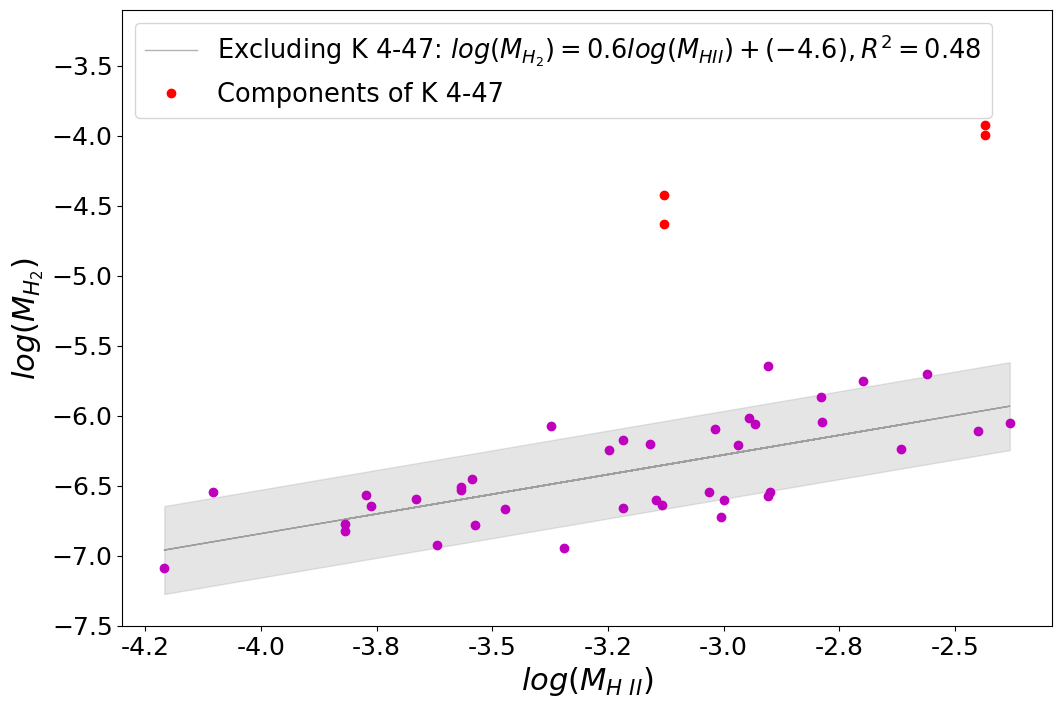}
    \caption{Correlation between ionized and warm molecular hydrogen mass for the LISs present in NGC~7662, NGC~7009 and NGC~6543 \citep{Akras2017,Akras2020}. The gray filled area represents the uncertainty of the regression line. LISs and outflows from K~4-47 were excluded from the linear fit due to their significant deviation from the bulk. The red points correspond to these nebular components, with the more massive ones representing the outflows.} 
    \label{fig:m_ion}
\end{figure}

The derivations of the above masses for a reasonable number of LISs allow the construction of the M$_{\rm H_2}$ versus M$_{\rm H~II}$ diagram, and also further explore the measurements 
from \citet{Akras2017,Akras2020}.
Nebular components from the PN K~4-47, marked with red dots in Fig.~\ref{fig:m_ion}, 
deviate significantly from the bulk of the data 
and are excluded from both the best-fit mass correlation and further discussion. 
K~4-47 
is a highly collimated PN with a pair of LISs located at the tips of its outflows. Employing photoionization and shock models, it was found that these LISs are
likely shock-heated, with shock velocities $\geq$150~km~s$^{-1}$ \citep{Corradi2000a,2004MNRAS.355...37G}. In the statistical analysis of LISs and host PNe carried out by \citet{2023MNRAS.525.1998M}, K~4-47  
frequently appear 
as an outlier 
-- in terms of electron temperatures, chemical abundances and various emission line ratios -- showing significant differences from the rest of the LISs.

A significant dispersion (R$^2\sim0.48$) is found in the log(M$_{H_2}$)--log(M$_{H{\sc II}})$ relation, which probably reflects the large uncertainties in the masses. The best-fit line is represented by:

\begin{dmath}\label{eq5}
    log\left(M_{H_{2}}\right)  = (0.6\pm0.1)~log\left(M_{HII}\right) - (4.6\pm0.3)
\end{dmath}
\vspace{0.2cm}

\noindent The median M$_{HII}$ for LISs in Table~\ref{tab:masses_akras}, excluding K~4-47, is $7.0\times10^{-4}$M$_{\odot}$. Using this value in Equation~\ref{eq5}, we obtain M$_{H_{2}}\sim 3.2\times10^{-7}$~M$_{\odot}$, while the median 
for this quantity is $3.0\times10^{-7}$M$_{\odot}$. 
According to \citet{2007MNRAS.382.1447M}, extinction affects the absolute intensity by less than 5\%, which does not impact our results or overall conclusions, given that the uncertainty in our mass estimates is even larger. Moreover, following \citet{1982ApJ...253..136S} and using Equations~\ref{eq1} and \ref{eq3} for the data in Table~\ref{tab:masses_akras}, we obtain:

\begin{dmath}\label{eq6}
    \centering
    \frac{M_{HII}}{M_{H_{2}}} = \frac{6.48\times10^{5}}{n_{e}}\frac{F_{Br\gamma}}{F_{H_{2}\lambda2.122}}
\end{dmath}

\noindent This corresponds to a difference of only 13 percent compared to Scoville's ratio (see their Eq.~8). Considering the $F_{Br\gamma}/F_{H_{2}}$ ratio, which ranges from $\sim$0.70 to 17.82, and the median electron density of 2325~cm$^{-3}$ for the LISs in the sample (excluding K~4-47), we find that their ionized masses (M$_{HII}$) are between approximately 200 to 5000 times larger than the warm molecular hydrogen masses (M$_{H_2}$).

\begin{table*}[ht!]
\caption{Positions, H$_2$~1-0-S(1) and Br$\gamma$ fluxes, warm molecular and ionized masses for various LISs in K~4-47 and NGC~7662 \citep{Akras2017}, as well as NGC~7009 and NGC~6543 \citep{Akras2020}, are presented. For PNe with multiple LISs, an average value of electron densities (n$_e$) was used, based on the published measurements for the LISs in each object. The n$_e$ were adopted from \citet{2004MNRAS.355...37G} for K~4-47, \citet{2009MNRAS.398.2166G} for NGC~7662, \citet{2003ApJ...597..975G} for NGC~7009 and \citet{1994ApJ...424..800B} for NGC~6543. }
\label{tab:masses_akras}
\centering
\small
\begin{tabular}{llllccccc}
\hline
Name	 & LISs	&  RA	          &  Dec	           &    Flux H$_2$~1-0-S(1)	    &  Flux Br$\gamma$    &  n$_e$[cm$^{-3}$]	 &  M$_{warm}$	     & M$_{ion}$     \\
         &      &                 &                    &   [$\times 10^{-15}$ erg s$^{-1}$cm$^{-2}$] & [$\times 10^{-15}$ erg s$^{-1}$cm$^{-2}$] &  & [$\times 10^{-07}$ M$_{\odot}$] & [$\times 10^{-04}$ M$_{\odot}$] \\
\hline
\multirow{4}{*}{K~4-47}	 & 1	    &  04:20:45.5    & +56:18:16.2    & 6.82  &	0.96  &	4600 &	375	 & 7.4      \\
         & 2	    &  04:20:44.9    & +56:18:10.6    &    4.29        &	0.96        &	4600 &	236	     & 7.4      \\
         & 3$^{\dag}$	    &  04:20:45.4    & +56:18:14.8    &    21.89       &	2.47        &	2400 &	1202     & 36.6      \\
         & 4$^{\dag}$	    &  04:20:45.1    & +56:18:12.4    &    18.38       &	2.47        &	2400 &	1009     & 36.6       \\
\hline
\multirow{29}{*}{NGC~7662}	 & 1  &  23:25:54.7    & +42:32:21.5    &    1.88  & 6.18 &	2325 &	 3.0  	 & 2.7      \\
         & 2	    &  23:25:54.5    & +42:32:21.6    &    1.96        &	6.18        &	2325 &	3.1  	 & 2.7      \\
         & 3	    &  23:25:54.6    & +42:32:17.5    &    1.05        &	6.62        &	2325 &	1.7  	 & 2.9      \\
         & 4	    &  23:25:54.6    & +42:32:16.2    &    4.23        &	13.80       &	2325 &	6.7  	 & 6.0      \\
         & 5	    &  23:25:54.4    & +42:32:14.8    &    5.14        &	21.90       &	2325 &	8.1  	 & 9.6      \\
         & 6	    &  23:25:54.7    & +42:32:15.4    &    1.38        &	7.68        &	2325 &	2.2  	 & 3.4      \\
         & 7	    &  23:25:54.8    & +42:32:13.8    &    8.63        &	37.10       &	2325 &	13.6   	 & 16.3      \\
         & 8	    &  23:25:54.9    & +42:32:09.7    &    12.60       &	62.80       &	2325 &	19.8   	 & 27.5      \\
         & 9	    &  23:25:54.9    & +42:32:09.1    &    3.95        &	24.50       &	2325 &	6.2  	 & 10.7      \\
         & 10	&  23:25:54.4    & +42:32:10.9    &    1.61        &	4.94        &	2325 &	2.5  	 & 2.2      \\
         & 11	&  23:25:54.6    & +42:32:09.6    &    1.44        &	3.94        &	2325 &	2.3  	 & 1.7      \\
         & 12	&  23:25:54.5    & +42:32:07.6    &    1.73        &	3.85        &	2325 &	2.7  	 & 1.7      \\
         & 13	&  23:25:54.7    & +42:32:06.7    &    1.81        &	21.20       &	2325 &	2.8  	 & 9.3      \\
         & 14	&  23:25:54.8    & +42:32:05.4    &    1.59        &	16.30       &	2325 &	2.5  	 & 7.1      \\
         & 15	&  23:25:54.9    & +42:32:05.5    &    1.39        &	13.80       &	2325 &	2.2  	 & 6.0      \\
         & 16	&  23:25:54.9    & +42:32:01.2    &    1.69        &	28.40       &	2325 &	2.7  	 & 12.4      \\
         & 17	&  23:25:54.7    & +42:32:00.3    &    1.59        &	22.90       &	2325 &	2.5  	 & 10.0      \\
         & 18	&  23:25:54.6    & +42:31:58.7    &    1.46        &	16.80       &	2325 &	2.3  	 & 7.4      \\
         & 19	&  23:25:53.7    & +42:31:58.2    &    5.62        &	94.70       &	2325 &	8.8  	 & 41.5       \\
         & 20	&  23:25:53.7    & +42:31:57.3    &    4.95        &	80.90       &	2325 &	7.8  	 & 35.5      \\
         & 21	&  23:25:53.7    & +42:31:52.0    &    14.30       &	28.40       &	2325 &	22.5  	 & 12.4      \\
         & 22	&  23:25:53.6    & +42:31:50.4    &    5.38        &	9.67        &	2325 &	8.5  	 & 4.2      \\
         & 23	&  23:25:53.5    & +42:31:48.2    &    0.96        &	3.47        &	2325 &	1.5  	 & 1.5       \\
         & 24	&  23:25:53.5    & +42:31:47.3    &    1.07        &	3.47        &	2325 &	1.7  	 & 1.5       \\
         & 25	&  23:25:53.2    & +42:32:02.9    &    5.72        &	37.20       &	2325 &	9.0  	 & 16.3      \\
         & 26	&  23:25:52.9    & +42:32:01.5    &    2.26        &	6.51        &	2325 &	3.6  	 & 2.9      \\
         & 27	&  23:25:52.9    & +42:31:57.6    &    11.20       &	45.70       &	2325 &	17.6  	 & 20.0      \\
         & 28	&  23:25:53.2    & +42:31:54.8    &    6.09        &	25.90       &	2325 &	9.6  	 & 11.4      \\
         & 29	&  23:25:54.2    & +42:32:18.8    &    0.72        &	10.30       &	2325 &	1.1  	 & 4.5      \\
\hline
\multirow{5}{*}{NGC~7009} & 1 &  21:04:12.5    & $-$11:21:41.9  & 3.71 &2.61        &	1650 &	2.9  	 & 0.8  \\
         & 2	    &  21:04:12.7    & $-$11:21:36.8    &    1.06        &	2.05        &	1650 &	0.8  	 & 0.6      \\
         & 3	    &  21:04:09.0    & $-$11:21:52.9    &    11.40       &	28.20       &	1200 &	8.8  	 & 11.7      \\
         & 4	    &  21:04:09.0    & $-$11:21:53.2    &    8.13        &	16.70       &	1200 &	6.2  	 & 6.9      \\
         & 5	    &  21:04:09.0    & $-$11:21:52.9    &    7.36        &	13.60       &	1200 &	5.7  	 & 5.6      \\
\hline
\multirow{4}{*}{NGC~6543} & 1 &  17:58:32.1 & +66:37:47.7 & 2.99    &	44.80        &	2200  &	 2.8  	 & 12.6  \\
         & 2	    &  17:58:32.1    & +66:37:47.4    &    6.03        &	86.30       &	2200 &	5.7  	 & 24.2      \\
         & 3	    &  17:58:32.2    & +66:37:46.9    &    1.26        &	8.56        &	2200 &	1.2  	 & 2.4      \\
         & 4	    &  17:58:33.8    & +66:38:11.7    &    1.97        &	35.10       &	2200 &	1.9  	 & 9.9     \\
\hline 
\end{tabular}
\tablefoot{$^{\dag}$Outflows from K~4-47 \citep{Akras2017}, although showing H$_2$ emission, are not considered LISs.}
\end{table*}

%--------------------------------------------------------------------
\section{Discussion}\label{sec5}
Since this work presents the first near-IR H$_2$~1-0 2.122~$\mu$m detections in the LISs of NGC~2392, NGC~6751, NGC~6818, NGC~6884 and NGC~7354, we begin by briefly characterizing these structures where warm molecular hydrogen was detected.

\vspace{0.2cm}
\noindent
(i) Multiple system of knotty and filamentary LISs, located at the inner part of the 
shell, in the high-excitation PN NGC~2392. It also displays high-velocity ([N~{\sc ii}]) jets directed toward the observer but not detected in H$_2$\citep{2012ApJ...761..172G,2021ApJ...909...44G}. The electron densities of the knots and filaments are 900-1000~cm$^{-3}$, 2 to 3 times lower that the electron densities in the nebular shell, while T$_{e}$ does not vary significantly \citep{1991ApJ...371..217B,2012ApJ...754...28Z}.

\vspace{0.2cm}
\noindent
(ii) Although NGC~6751 displays a large halo in which H$_2$ is also detected \citep{2010ApJ...722.1260C}, the clumpy LISs of its inner regions, including the jet-like structures \citep[ansae, in][]{1991ApJ...376..150C} are the main focus here. T$_{e}$ has been found to be similar across the different nebular regions. On the other hand, the jet-like structures, knots and rim have, respectively, $\sim$200~cm$^{-3}$, $\sim$1600~cm$^{-3}$ and $\sim$2500~cm$^{-3}$ \citep{1991ApJ...376..150C}.

\vspace{0.2cm}
\noindent
(iii) NGC~6818 has multiple [N~{\sc ii}]-bright filamentary structures distributed throughout the nebula \citep["moustaches",][]{2003A&A...400..161B}, which are also observed in H$_2$. These structures are peculiar due to their higher kinematics gradients \citep{1999ApJ...514..878H} and electron densities compared to the other nebular components. Across the nebula, N$_e$[S{\sc~ii}] varies between 1500 and 2800~cm$^{-3}$, with the equatorial filamentary structures being the densest regions \citep[][]{2003A&A...400..161B,2003MNRAS.345..186T, 2005A&A...436..953P}.

\vspace{0.2cm}
\noindent
(iv) [N~{\sc ii}] HST images and H$_2$ NIRI images have revealed that the LISs of NGC 6884 form a pair of knotty, arc-like structures \citep{2002AJ....123.2666P}.
T$_e$ show a slight variation between the low- ($\sim$8500~K) and moderate-ionization ($\sim$10000~K) regions of the nebula. The low-ionization regions have N$_e$[O~{\sc ii}]$ of \sim$16000~cm$^{-3}$ while, based on [A~{\sc iv}] and [Cl~{\sc iii}] diagnostics, the rest of the nebula has values of $\sim6000~cm^{-3}$ \citep{1997ApJS..108..503H}.

\vspace{0.2cm}
\noindent
(v) The equatorial knots and jet-like structures of NGC~7354, detected in H$_2$, were studied by \citet{1997ApJ...487..304H,2010AJ....139.1426C}, who derived their temperatures and densities. 
The pair of jet-like structures, equatorial knots, and high-ionization shell are characterized by N$_e$ (T$_e$) values of $\sim$970~cm$^{-3}$ (10000-12000~K), $\sim$3000~cm$^{-3}$ (10000-13000~K), 1500-2600~cm$^{-3}$ ($\sim$10000-13000~K), respectively.

\vspace{0.3cm}
Regardless of their physical conditions, the nebular components or LISs, where H$_2$ emission is unambiguously detected, 
are associated with structures 
that are prominent in the [{\sc N~ii}] emission line. In most cases, the H$_2$ detections are associated with structures that, relative to the rims and shells of their host nebulae, have lower electron densities. 
Thus, by doubling the sample of known PNe with LISs detected in H$_2$ line, we verify that LISs are indeed composed of partially ionized and excited molecular gas. In addition to the LISs, which were the main focus of this narrowband survey, H$_2$ emission was also detected for the first time in the halos of NGC~6751 and NGC~6884.

Moreover, for the first time, the molecular mass of excited (\textit{warm}) H$_2$ gas in LISs was estimated for our sample of five PNe, along with the LISs in four PNe from the literature. The corresponding ionized masses of these LISs were also calculated. Based on the range of near-IR H$_2$ and Br$\gamma$ fluxes reported by \citet{Akras2017,Akras2020}, 
we found that the ionized mass in LISs is between 200 and 5000 times greater than the excited molecular mass. 

As mentioned in Section~\ref{sec4}, our data do not represent the distribution of the cold molecular component, but only the warm component. 
Estimating the H$_2$ mass of the cold counterpart is commonly done through CO emission. However, this method requires converting the CO column density into H${_2}$ mass, and for this, it is necessary to assume a conversion factor for the CO/H${_2}$ ratio --~a highly uncertain quantity \citep[see, e.g., ][]{2013ARA&A..51..207B}.

No actual measurements of the mass of the cold components are available for the LISs discussed in the present work due to the lack of any CO detection so far. 
\citet{2020MNRAS.491..758A} analyzed ALMA CO observations of a globule in the Helix nebula, specifically the C1 globule previously detected in CO and H$_2$ by \citet{2002ApJ...573L..55H}. Using CO emission lines as H$_2$ tracers, the molecular mass of the globule's head has been estimated $(0.8-1.3)\times$10$^{-5}$M$_{\odot}$ \citep{2020MNRAS.491..758A} and $\ \gtrapprox1\times$10$^{-5}$M$_{\odot}$ \citep{2002ApJ...573L..55H}, assuming a CO/H$_2$ ratio of $\sim$10$^{-4}$ \citet{1992ApJ...401L..43H,2002ApJ...573L..55H}. 
Considering the H$_2$~2.122~$\mu$m flux reported by \citet{2002ApJ...573L..55H} at the globule's head (whereas the tail was significantly fainter in this emission), we estimated the warm molecular mass using Equation~\ref{eq2} to be $\sim1.5\times10^{-8}$~M$_{\odot}$. This suggests that the cold H$_2$ mass in the C1 head is approximately $5.3-8.7\times10^{2}$ times greater than the warm H$_2$ mass, yielding a mean ratio of $M_{cold}/M_{warm}\sim7\times10^{2}$. Based on this, if a cold H$_2$ gas component is present in our sample of LISs, their masses should range between $(0.3-7.3)\times10^{-4}$~M$_{\odot}$, with a mean value of 3.8$\times$10$^{-4}$~M$_{\odot}$.  This is approximately half of the median ionized mass found for LISs in Section~\ref{sec4.1}.

Since no other LISs have been detected in both H$_2$ and CO so far, direct comparisons are not possible. Nevertheless, the cold H$_2$ mass in the nebula NGC~3132 has also been computed based on Submillimeter Array (SMA) observations, ranging between $\sim$0.015 and $\sim$0.15~M$_\odot$, assuming a CO/H${_2}$ conversion factor between $10^{-4}$ and $10^{-5}$ \citep{2024ApJ...965...21K}.

The presence of molecular gas in LISs offers new insight to better understand their origin and nature. The new detections of H$_2$ emission from LISs 
have doubled the total number of PNe (nine in total) with LISs directly associated with warm molecular gas. This provides further support for the scenario in which LISs may represent mini-photodissociation regions (PDRs) illuminated by the hard UV radiation from the central stars of PNe. Nevertheless, it should be noted that the potential PDRs at the surfaces of LISs may differ from the 
PDRs typically found in high-mass star-forming regions \citep{1997ARA&A..35..179H}. The very low ionization parameters of PNe' LISs  may cause the ionization and dissociation fronts of H$_2$ to merge \citep{Bertoldi1996,Henney2007}.

The survival of molecular
gas in these nebular structures raises an important question:
\textit{how %has it survived its 
does it survive the dissociation due to the energetic UV photons of the central star?} Recent detection of the [C~{\sc i}]~$\lambda$8727 emission line from the LISs of NGC~7009 \citep{2024A&A...689A..14A} and NGC~3242 \citep{Konstantinou2025} suggests the photoevaporation of the molecular gas, which becomes dissociated and ionized as it flows away from the knots. \citet{2024A&A...689A..14A} demonstrated, at least for the northeastern LIS of NGC~7009, a stratified ionization structure at the interface between the outer ionized gas and the dense molecular core (i.e. within a transition zone). The emission profiles of that LIS show [C~{\sc i}] arising between [N~{\sc ii}] and H$_2$, which contrasts with what is observed in PDRs of star-forming regions \citep[e.g.][]{2021MNRAS.502.4597H}.

As mentioned before, the physical conditions in LISs may deviate significantly from those of typical high-mass star-forming PDRs. Nevertheless,
LISs appear to preserve some key features of 
typical PDRs, such as a stratified emission region, high total densities, and low temperatures. 
\citet[][]{1997ARA&A..35..179H} depicted a schematic of a PDR structure, where the {\sc H~ii} component is followed by a thin {\sc H~ii}/{\sc H~i} layer that absorbs Lyman-continuum photons. These layers are followed by large columns densities of {\sc O}, {\sc C}, {\sc C$+$}, {\sc CO} and vibrationally excited H$_2$. The typical total densities (10$^{3-5}$~cm$^{-3}$) and temperatures (between 200 and 1000~K) of PDRs are comparable to those mentioned by \citet{2020ApJ...889...13B}, where the authors have predicted high densities (10$^{6-7}$~cm$^{-3}$) and low temperatures (3$\times$10$^{0-3}$~K) in the interior of LISs. 
Such high densities has also been suggested for the LISs in NGC~6720 and NGC~3132 \citep{2024MNRAS.528.3392W,2022NatAs...6.1421D}.

\section{Conclusion}\label{sec6}
In this study we presented a deep narrowband H$_2$ imaging 
study of five PNe with several LISs embedded. The number of PNe with LISs detected in the near-IR H$_2$~2.122~$\mu$m ro-vibrational line has doubled (to 9). The conclusions of the study are as follow:
\begin{itemize}
   \item Although the detected emission corresponds to warm H$_2$, it directly implies the presence of a colder molecular component. The coexistence of warm and cold H$_2$ supports the model expectations that LISs contain significant amounts of molecular material, helping reconcile the low \textit{electron} densities derived from the ionized gas with models that predict total densities in LISs to be orders of magnitude higher than in the host nebula;
  \item The warm H$_2$ masses of the LISs studied in this work range from 0.4 to 10$\times$10$^{-7}$~M$_{\odot}$, with an average of 4.6$\times$10$^{-7}$~M$_{\odot}$; 
  \item The ionized masses (M$_{HII}$) of the LISs observed in Br$\gamma$ vary between 0.6 and 41.5$\times$10$^{-4}$~M$_{\odot}$, with an average value of 7$\times$10$^{-4}$~M$_{\odot}$; 
  \item We found that the excited H$_2$ molecular mass (M$_{H_2}$) in LISs is between 200 and 5000 times lower than the corresponding ionized mass.
\end{itemize}

\begin{acknowledgements}
We thank William Henney, the referee, for his comments and suggestions. We would like to thank Mateus Dias Ribeiro for his essential help with techniques to degrade the images to improve the continuum subtraction and the so reliable result. This research project was partially supported by the Consejo Nacional de Investigaciones Científicas y Técnicas (CONICET, Argentina). MBM was supported by a CAPES (The Brazilian Federal Agency for Support and Evaluation of Graduate Education within the Education Ministry) fellowship at the beginning
of this study. SA acknowledges the research project implemented in the frame-work of H.F.R.I call \lq\lq Basic research financing (Horizontal support of all Sciences) \rq\rq~under the National Recovery and Resilience Plan \lq\lq Greece 2.0\rq\rq~funded by the European Union -- NextGenerationEU (H.F.R.I. Project Number: 15665). DRG acknowledges FAPERJ (E-26/211.527/2023) and CNPq (315307/2023-4) for partical support.
\\
This research is based on observations acquired through the Gemini Observatory Archive at NSF NOIRLab and processed using DRAGONS (Data Reduction for Astronomy from Gemini Observatory North and South). This work makes use of Hubble Space Telescope data obtained from the Hubble Legacy Archive (STScI/NASA, ST-ECF/ESA, CADC/NRC/CSA).

\end{acknowledgements}

\bibliographystyle{aa} % style aa.bst
\bibliography{refes} % your references Yourfile.bib

\end{document}